\documentclass{aa}
\usepackage{graphicx,amssymb}
\usepackage{astron}
\begin{document}

\title{$H_0$ Measurement from  VLT Deep I--band Surface Brightness 
Fluctuations in NGC~564 and NGC~7619
\thanks{Based on observations performed at the European 
Southern Observatory, Paranal, Chile ESO program N$^o$~66.A-0361}}

\titlerunning{I-band SBF in NGC~564 and NGC~7619}

  \author{S. Mei \inst{1}, M. Scodeggio \inst{2}, D.R. Silva \inst{3}, 
          P.J. Quinn \inst{3}
     }

\authorrunning{Mei, Scodeggio, Silva $\&$ Quinn}

   \offprints{S. Mei; simona.mei@ias.fr}

   \institute{Institut d'Astrophysique Spatiale, 
                             Universit\'e Paris-Sud, B\^atiment 121
                             F-91405 Orsay, France
\and    
Istituto di Fisica Cosmica G. Occhialini, Via Bassini 15, 20133
Milano, Italy
\and
European Southern Observatory, Karl-Schwarzschild-Strasse 2, 85748 
Garching, Germany 
         }

   \date{Received June 28, 2002; }

   \abstract{We have measured the Hubble constant $H_0$ in NGC~564 at
$cz \approx 5800$ km~s$^{-1}$ and in NGC~7619 at $cz \approx 3700$ km
s$^{-1}$ with deep I--band Surface Brightness Fluctuation distance
measurements at the ESO Very Large Telescope (VLT). We obtain~$H_0 =
70 \pm 7 \pm 6$ km~s$^{-1}$/Mpc for NGC~564 and $H_0 = 68 \pm 6 \pm 6$
for NGC~7619. The actual SBF sample used for the measurement of $H_0$
in the Hubble Space Telescope Key Project on the Extragalactic
Distance Scale (Freedman et al. 2001) amounts to six galaxies.  When
we combine the measurements from this work with our previous VLT
I--band SBF distance measurement in IC~4296 (Mei et al. 2000), we
obtain~:~$H_0$ = 68 $\pm$ 5 $\pm$ 6 km~s$^{-1}$/Mpc. When we add the
Freedman et al. (2001) SBF sample, we obtain $H_0 = 71 \pm 4 \pm 6$ km
s$^{-1}$/Mpc.

   }

   \maketitle
%

\section{Introduction}

Future Cosmic Microwave Background missions like MAP 
(http://map.gsfc.nasa.gov/)
and Planck (http://astro.estec.esa.nl/SA-general/Projects/Planck/)
promise to obtain definitive measurements for cosmological
parameters, achieving an accuracy around $1\%$. Over the last few 
years, however, the accuracy of such measurements has improved 
considerably, as many efforts have been carried out to obtain 
accurate measurements for the Hubble constant $H_0$, to
reduce the uncertainty in its determination from 50$\%$ to approximately
$10\%$. The most important among these efforts has been the Hubble Space
Telescope Key Project on the Extragalactic Distance Scale (HSTKP;
Freedman et al. 2001 and references therein) where the value of $H_0$
was measured to be $H_0 = 72 \pm 8$  km~s$^{-1}$, based on a coherent
Cepheid calibration of four of the most dependable and precise methods
for galaxy distance measurements: Tully--Fisher, Fundamental Plane,
SNIa and Surface Brightness Fluctuations. Other important
contributions were provided, among others, by Ajhar et al. (2001);
Blakeslee et al. (2001); Jensen et al. (2001); Gibson \& Stetson
(2001); Liu \& Graham (2001); Tonry et al. (2000);  Suntzeff et 
al. (1999); Jha et al. (1999); Giovanelli et al. (1997); Hjorth 
\& Tanvir (1997).

$H_0$ measurements are affected not only by uncertainties in the 
derivation of the cosmological distance ladder, but also by the 
presence of local deviations from the smooth Hubble flow. 
The average peculiar velocity for galaxies and clusters of galaxies 
are of the order of 100 - 300~km~s$^{-1}$ (Davis et al. 1997; 
Giovanelli et al. 1998; Dale et al. 1999).
Therefore to minimise their effect on the determination of $H_0$,
it is preferable to select target objects beyond 4000~ km~s$^{-1}$,
where the peculiar motions amount to no more than $\approx 10 \% - 15
\% $ of the total velocity. This requirement has led to a number
of projects trying to build samples of redshift independent distance
estimates restricted to galaxies located beyond 4000 - 5000~km~s$^{-1}$
(Dale et al. 1999; Hudson et al. 2001; Colless et al. 2001; Jensen et
al. 2001).

Among the methods used to obtain accurate galaxy distance estimates,
Surface Brightness Fluctuations (SBF) are, at present, one of the most
accurate and better physically understood methods available.  The
method was introduced by Tonry \& Schneider 1988 (a recent review has
been given by Blakeslee et al. 1999a) and is based on the very simple
fact that the Poissonian distribution of unresolved stars in a galaxy
produces fluctuations in each pixel of the galaxy image.  The variance
of these fluctuations is inversely proportional to the square of the
galaxy distance. The SBF amplitude is defined to be precisely this
variance, normalised to the mean flux of the galaxy in each pixel
\cite{ts88}. 


\begin{table*}[!htf]

\begin{flushleft}
\begin{tabular} {|c|c|c|c|c|c|c|c|c|} \hline 
Name&RA &DEC &Type&$m_V^1$&$V-I^1$&$cz_{LG}^2 $&$cz_{CMB}^2$&$V^{pec 2}_{CMB}$ \\ \hline
&(2000)&(2000)&&mag&mag&km s$^{-1}$& km s$^{-1}$&km s$^{-1}$ \\ \hline
NGC 564&01 27 48.31 & -01 52 47.68 &E&13.01&1.2&5843& 5037&-52 \\ \hline
NGC 7619&23 20 14.68 &+08 12 23.3 &E&11.90&1.23&3758&3519&-256 \\ \hline
\end{tabular}

\end{flushleft}

$^1$ Prugniel $\&$ Heraudeau 1998, Poulain $\&$ Nieto 1994 \\
$^2$ Kelson et al. 2000 and references therein, Giovanelli 1998, Dale 1999

\caption{General properties of NGC 564 and NGC 7619, from the web archive Simbad (http://simbad.u-strasbg.fr) and the NASA/IPAC Extragalactic Database (NED) (http://nedwww.ipac.caltech.edu/)} \label{tab-vlt}
\end{table*}
                 

\begin{table*}[!htf]

\begin{flushleft}
\begin{tabular} {|c|c|c|} \hline 
Port&FORS1 Ron ($e^{-}$)&Gain ($e^{-}$/ADU)\\ \hline

A&5.01$\pm$0.16&1.48$\pm$0.04 \\
B&5.42$\pm$0.17&1.78$\pm$0.05 \\
C&5.18$\pm$0.16&1.56$\pm$0.04 \\
D&5.03$\pm$0.16&1.63$\pm$0.05 \\ \hline

\end{tabular}

\end{flushleft}

\caption{Read--out Noise and Gain for the four ports of FORS1 on VLT UT1} \label{tab-fors}
\end{table*}


The absolute magnitude of the fluctuation is not a
constant, but depends on the age and metallicity of the stellar
population within the galaxy. Tonry et al. (1997) and Tonry et
al. (2001) have quantified these dependencies using an extensive
sample of I and V band observations, which are used to empirically
calibrate the dependence of the I -- band fluctuation magnitude on (V -- I)
colour. The absolute calibration is then obtained using a set of 5
galaxies with independent Cepheid distances.

As part of the HSTKP, Ferrarese et al. (2000a) have obtained their own
calibration based on accurate Cepheid distances to six spiral galaxies
with SBF measurements within 1200~km~s$^{-1}$. The Cepheid distances
were part of a larger dataset used by HSTKP. The two calibrations are
consistent within the errors.

Alternatively, theoretical calibrations have been obtained starting
from synthesis models of the galaxy stellar population (Ajhar et
al. 2001; Liu et al. 2000; Blakeslee et al. 2001). From Ajhar et
al. (2001) the zero points of these calibrations provide an average
value of the Hubble constant ($H_0 = 77 \pm 7$~km~s$^{-1}$/Mpc) that
is in good agreement with the observational results.

With HST and 8-m class telescopes SBF measurements have been extended
beyond the local universe ($cz < 4000$~km~s$^{-1}$). While the survey
by Tonry et al. (2001) was limited to galaxies with $cz < 4000$~km
s$^{-1}$, using HST observations it has been possible to determine
distances for more distant galaxies (Thomsen et al. 1997; Lauer et
al. 1998; Pahre et al. 1999; Jensen et al. 2001; Liu et al. 2001;
Ajhar et al. 2001), up to approximately $10,000$~km~s$^{-1}$.

In this paper we extend our ground-based I-band SBF measurements to
beyond $\approx 5000$~km~s$^{-1}$. Advantages and disadvantages of I-
versus K-band ground based SBF observations were discussed using
theoretical simulations in Mei et al. (2001).  Deep I--band SBF
observations were obtained for two galaxies : NGC~564, in Abell 194,
at $cz \approx 5800$~km~s$^{-1}$, and NGC~7619 in the Pegasus cluster
at $cz \approx 3700$~km~s$^{-1}$.  Both Abell 194 and Pegasus are
fiducial clusters in the derivation of observational templates for the
Tully--Fisher (TF) and Fundamental Plane (FP) relations (see for
example Mould et al. 1991, 1993; Jorgensen et al. 1996; Giovanelli et
al. 1998; Scodeggio et al. 1998; Dale et al. 1999; Hudson et
al. 2001), while Abell 194 is also part of the sample of clusters
studied by Lauer \& Postman (1992, 1994). Although distance estimates
obtained with the TF and FP relations are comparatively less accurate
than those obtained with the SBF method, they are relatively easier to
obtain, even for objects located at larger distances than those
sampled so far by SBF observations. Therefore the most complete and
farthest reaching mappings of the local universe currently make use of
TF- and FP-based distances, and they will likely do so in the near
future. One of the main motivations behind the selection of targets
for this work was the possibility of providing a bridge between SBF
distance estimates and TF or FP results, as this is the most direct
way to improve the accuracy of the calibration for the latter
relations, and consequently that of all distance estimates based on
these relations. This effort extends to larger distances the matching
of SBF and FP distance scales recently presented by Blakeslee et
al. (2002).

In $\S$ 2 we describe our observations, and the SBF
analysis.  The $H_0$ measurements and a summary of this analysis are
presented in $\S$ 3 and 4.


\begin{table*}[!htf]

\begin{flushleft}
\begin{tabular} {ccccccc} \hline 

Galaxy&Band&Date&Frames&Exp.Time (sec)\\ \hline
NGC 7619&V& 3 Nov 00&5&100\\ 
&I&5 Oct 00&4&60 \\ 
&I&5 Oct 00&55&150 \\ 
NGC 564&V&19 Aug 01&5& 100\\ 

&I&3 Nov 00&81&65\\ 
&I&2 Dec 00&53&65\\ 
&I&19 Dec 00&41&65 \\ 
&I&22 Dec 00&41&65 \\ 
&I&18 Jan 01&41&65 \\ 
&I&21 Jan 01&36&65\\ 
 \hline
 
\end{tabular}

\end{flushleft}
\caption{Observations} \label{tab-obs}
\end{table*}


\section{Surface Brightness Fluctuation analysis}

\subsection{Observations}

NGC~564 and NGC~7619 were observed in service mode at the Very Large
Telescope unit UT1 (Antu) at the European Southern Observatory in
Paranal, Chile, using the FORS1 (FOcal Reducer and low dispersion
Spectrograph) imaging camera with a 2048 x 2048 pixel Tektronix
CCD. The properties of these galaxies are summarised in
Table~\ref{tab-vlt}.

The FORS1 high gain, standard resolution mode was used, with a
pixel scale 0.2$\arcsec$/pixel and field of view 6.8$\arcmin$ x
6.8$\arcmin$.  The detector had four ports with different read out
noise and gain. We show the gains for the different ports in
Table~\ref{tab-fors}. The dark current is negligible
($<2e^-$/pixel/hour at nominal operating temperature).

I--band and V--band data were obtained, to have an accurate
V--I colour to input into the Tonry et al. (2000) calibration of I--band SBF
fluctuations.  For NGC~564, the nights of observation were on 5
October, 3 November, 2, 19, 22 December 2000, and 1, 18 and 21 January 2001
for the I--band and on 19 August 2001 for the V--band.
The raw data were 293 galaxy exposures each of $\approx$~65 seconds in the I--band and 5 galaxy exposures each of 100 seconds in the V--band. The total observation time was 5.3 hrs in the I--band and 500 sec in the V--band.
  For
NGC~7619, the nights of observation were on 5 October 2000 for the I--band
and the 3 November 2000 for the V--band. 
The raw data were 55 galaxy exposures each of $\approx$~150 seconds, and 4 exposures of 60 seconds in the I--band, and 5 galaxy exposures each of 100 seconds in the V--band. The total observation time was 2.4 hrs in the I--band and 500 sec in the V--band.
A summary of observations is in Table~\ref{tab-obs}.  

The seeing was
$\approx 0.8\arcsec$ FWHM on the nights of observations. 
Data observed on different nights were calibrated to one 
photometric night observations for each galaxy in each band,
 by the use of reference stars in the images. In the case of NGC~7619,
some of the images in I--band were not photometric and the final added
image has been calibrated on a combination of photometric images.

All observations were obtained in VLT Service Mode.  Bias corrected,
flat-fielded, and trimmed frames were delivered by ESO and used by us
without further processing.  Further details about pipeline processed
FORS science products and calibration can be found on the ESO Quality 
Control Web pages (http://www.eso.org/qc/).
  
The photometric zero-point and the extinction coefficient were
extracted from observations of Landolt standards \cite{lan92} 
and confirmed using the historical data
     available from the ESO Quality Control Web pages.
 The I--band filter in the FORS1 camera is a Bessel I filter, but all
measurements were transformed into Kron-Cousins I--band magnitudes. In
this magnitude scale, for NGC~564 the photometric zero-point is
$m_1=26.59 \pm 0.02$ mag ($m_{1}$ is the zero magnitude which corresponds 
to a flux of 1 ADUs$^{-1}$) and the extinction coefficient $0.06 \pm
0.01$ mag. The colour term correction in (V--I) is --0.05 $\pm$ 0.03
mag.  For NGC~7619, $m_1=26.60 \pm 0.02$ mag and the extinction
coefficient $0.07 \pm 0.01$ mag. 

In the V--band, the photometric
zero-point was $m_1=27.53 \pm 0.02$ mag for NGC~564 and 
 $m_1=27.45 \pm 0.02$ for NGC~7619, and the extinction coefficient
was $0.11 \pm 0.05$ mag for NGC~564 and $0.12 \pm 0.05$ for
NGC~7619. The colour term correction in (V--I) is 0.03 $\pm$ 0.03 mag.

Bad pixels and cosmic rays were eliminated by a sigma clipping algorithm
while combining the images using the IRAF \footnote{The Image Reduction
 and Analysis Facility (IRAF) is distributed by the National Optical
Astronomy Observatories}  task IMCOMBINE. 
The images have been scaled in exposure time and median value.
Sub-pixel
registration was not used to avoid the introducing correlated noise
between the pixels in the images.


\begin{figure}[!htf]

 \centering
\includegraphics[width=7cm]{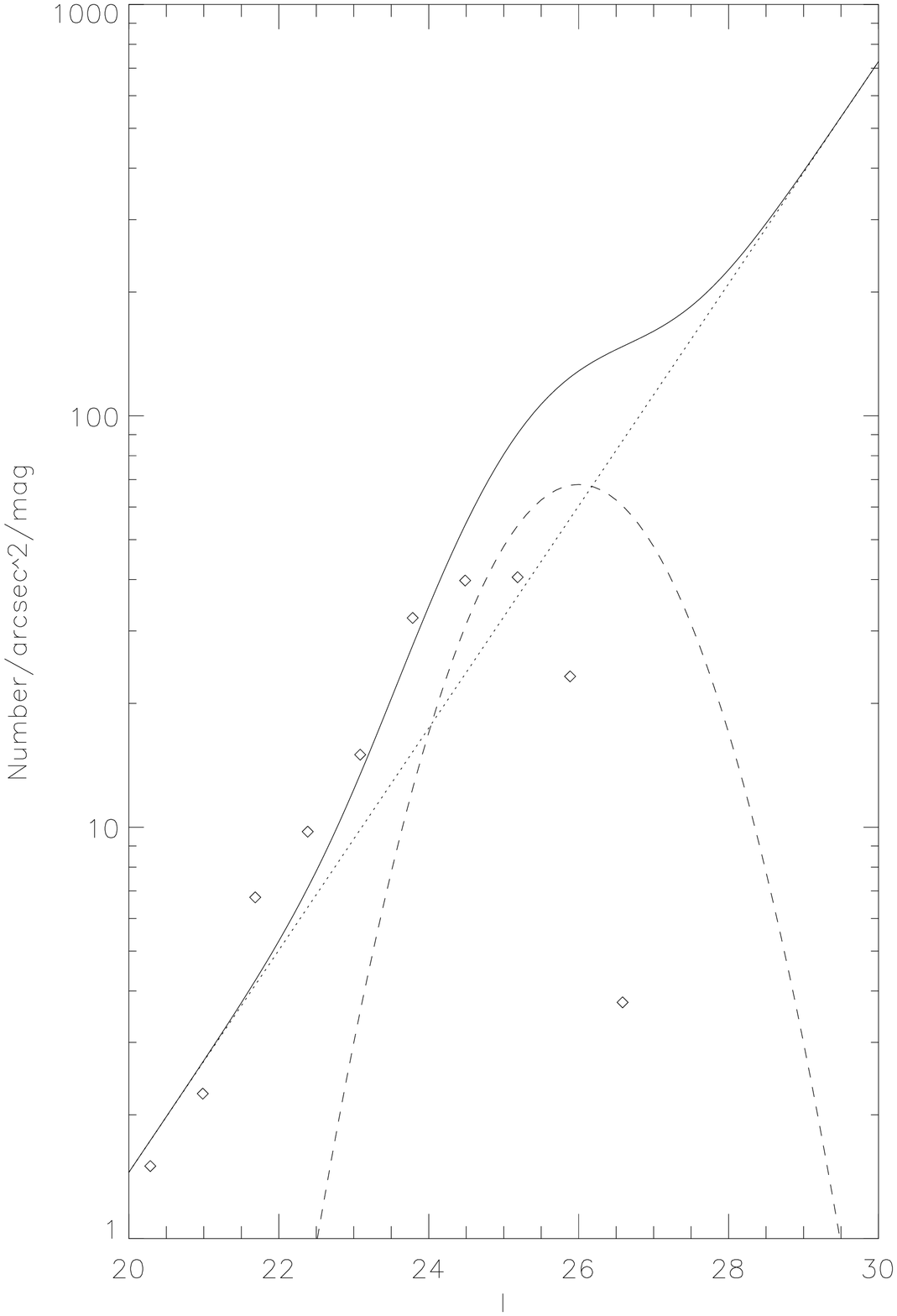}

\caption{We show NGC 564 external source luminosity function between
14$\arcsec$ and 23$\arcsec$ from the center of the galaxy. The solid line shows the sum of the
globular cluster luminosity function plus the background galaxy
luminosity function. The dashed line shows the fitted globular cluster
luminosity function, the dotted line the background galaxy luminosity
function. } \label{fig-564lum}

\end{figure}



\begin{figure}[!htf]

 \centering
\includegraphics[width=7cm]{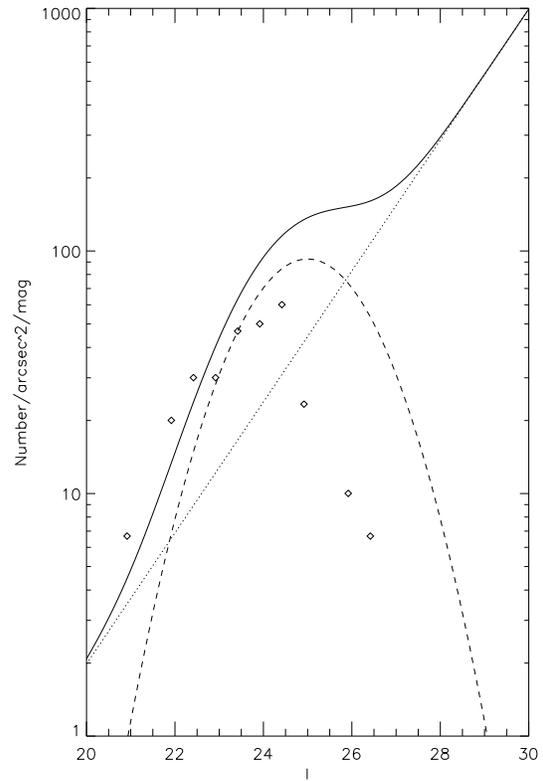}

\caption{We show NGC 7619 external source luminosity function between
20$\arcsec$ and 52$\arcsec$ from the center of the galaxy. The solid line shows the sum of the
globular cluster luminosity function plus the background galaxy
luminosity function. The dashed line shows the fitted globular cluster
luminosity function, the dotted line the background galaxy luminosity
function. } \label{fig-7619lum}

\end{figure}


\subsection[SBF Measurements]{SBF Measurements}

The data were analysed with a standard two--step SBF extraction 
technique, as detailed, for example, by Tonry \& Schneider 1988. 
Firstly, a smooth galaxy profile was subtracted from the 
image. The smoothed image was derived by
fitting galaxy isophotes to the original image with the IRAF task
ISOPHOTE, once visible external sources were subtracted.  
To account for
residual errors that can be due to the galaxy profile subtraction, 
the resulting image was then smoothed on a scale ten times the
width of the PSF, once additional point sources were identified 
and subtracted using the software tool Sextractor \cite{ber96}, 
and subtracted from the original image.

Secondly, a new galaxy model was fit and subtracted to this difference image.
In such way, the sky is also subtracted.
In order to make the SBF fluctuations constant
across the image, the difference image was finally divided by the 
square root of the new galaxy model.

This final image was then divided into different annuli, in each of
which the power spectrum of the image fluctuations was calculated. The
resulting power spectrum was azimuthally averaged, and normalised to
the number of non-zero points in the annulus.
The external point sources brighter than a magnitude $m_{cut}$ where
the completeness function of each annulus was greater than 50$\%$ were
masked out.

Two components contribute to the total image power spectrum:
the constant power spectrum due to the white noise, $P_1$, and the
power spectrum of the fluctuations and point sources that are both
convolved by the PSF in the spatial domain. In the Fourier domain
these latter terms are given by a constant, $P_0$, multiplied by the
power spectrum of the PSF:
\begin{equation}
E_{gal}= P_0 \  E_{PSF} +P_1 .
\end{equation}


\begin{figure}[!htf]

 \centering
\includegraphics[width=8cm]{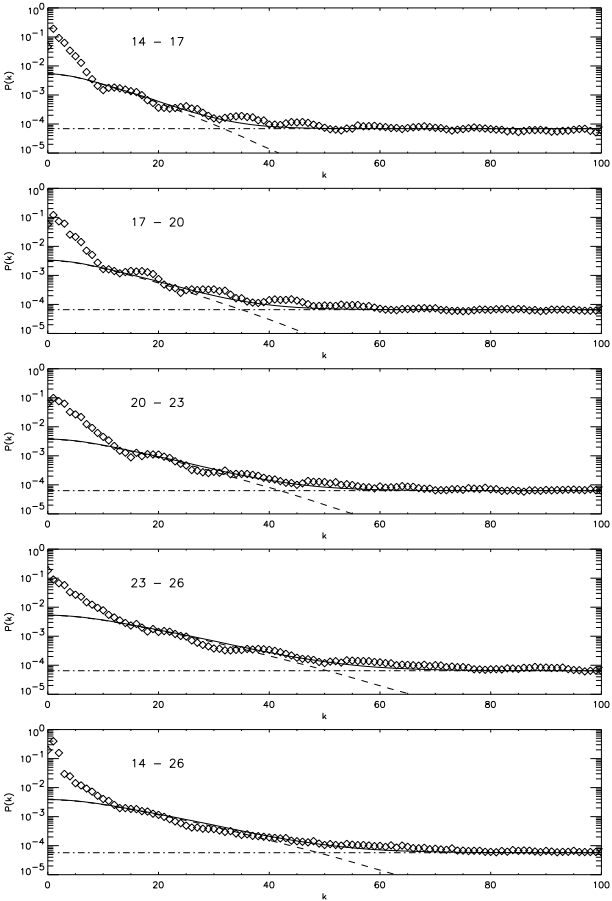}

\caption{NGC 564 SBF measurement. We show NGC~564 power spectrum as was fitted in
five different annuli of width $\approx$ 3 $\arcsec$ in the galaxy and
on the full field up to 26$\arcsec$. The fit of the power spectrum is
given by the solid line, the PSF power spectrum by the dashed line
and the dashed-dotted line is the fitted constant white noise
spectrum. } \label{fig-564spe}

\end{figure}


\begin{table*}[!htf]

\begin{flushleft}
\begin{tabular} {|c|c|c|c|c|c|c|c|c|c|c|c|c|c|} \hline 

Annulus ('') & P$_0$ & $\sigma_{P_0}$ &$P_{es}$&$\sigma_{P_es}$& \small (P$_0$ -- P$_{es}$)/P$_1$&$(V-I)_o$ &$\overline M_I$&$\overline I_{o,k} $& $\sigma_{\overline I_{o,k}}$&$\overline I_{o,k} - \overline M_I$&$\sigma_{(\overline I_{o,k} - \overline M_I)}$\\ \hline

14 - 17&0.0071 & 0.001&0.003  &0.001&68&1.20&-1.52&32.4&0.5 &33.92&0.5\\ \hline
17 - 20&0.0043 &  0.001&0.0015 &0.0005&46&1.20&-1.52&32.8&0.4 &34.32&0.4\\ \hline
20 - 23&0.0034 & 0.0004&0.0015 &0.0005&32&1.21&-1.47&33.2& 0.3  &34.67&0.4\\ \hline
23 - 26&0.0040 &  0.0005&0.0015 &0.0005&42&1.22&-1.43&32.9&0.3  &34.33& 0.4\\ \hline
14 - 26&0.0032  &0.0002& 0.0015 &0.0005&28&1.20&-1.52&33.2  &0.3  &34.72 & 0.4 \\ \hline

 \hline
 Mean  All& -- &-- & --&--&--&--&--&--&--&34.39 &0.18  \\ \hline

\end{tabular}

\end{flushleft}
\caption{SBF measurements for various annuli of NGC 564} \label{tab-564annu}
\end{table*}


To compute $P_0$ and $P_1$, a robust linear least squares fit was
made by minimising absolute deviation (Numerical Recipes, Press et
al. 1992). As part of the fit, to calculate $E_{PSF}$ a point spread
function (PSF) profile was determined from the bright stars in the
image and normalised to 1 ADUs$^{-1}$, following Tonry and Schneider (1988). 
This means that $P_0$ is directly the power that we wish to measure.

Low wave number ($k$) points were
excluded from the fit (pixel scale greater than 20), because they
are contaminated by the galaxy subtraction and subsequent smoothing.

As stated above, $P_0$ contains contributions from both SBF and
unresolved point sources.  The point source contribution $P_{es}$ was
estimated from the equations:
\begin{equation}
P_{es}=\sigma^2_{gc}+\sigma^2_{bg}
\end{equation}
following, i.e., Blakeslee \& Tonry (1995); $\sigma^2_{gc}$ is the
contribution to the fluctuations given by globular clusters,
$\sigma^2_{bg}$ is the contribution by background galaxies.  We
adopted the following luminosity function for the globular clusters:
\begin{equation}
N_{gc}(m) = \frac{N_{ogc}}{\sqrt{2\pi}\sigma} 
e^{\frac{-(m-m^{gc}_{peak})^2}{2\sigma^2}}
\end{equation}
 We have assumed as initial $M^{gc}_{peakI} \approx -8.5$ and
$\sigma=1.35$ \cite{fer00b} with standard globular cluster colours
\cite{geb00}.  For NGC~564, we have fitted to our detected globular
cluster number counts an initial maximum likelihood
$m^{gc}_{peakI}=26$ and have assumed an initial galaxy distance
modulus 34.4, from Kelson et al. (2000) and references therein.  For
NGC~7619, we have fitted to our detected globular cluster number
counts an initial maximum likelihood $m^{gc}_{peakI}=25$ and have
assumed an initial galaxy distance modulus 33.5, from Dale et
al. (1999).  These initial values were iterated in the process of the
SBF distance modulus calculation. 
The error on the external source contribution includes the uncertainties 
due to this iteration process. This last error has been calculated as 
the standard deviation of the estimation of the external source contribution 
for all the iterations, till stabilisation.

 For the background galaxies, a
power-law luminosity function was used:
\begin{equation}
  N_{bg}(m) = N_{obg} 10^{\gamma m} 
\end{equation}
 with $\gamma = 0.27$ \cite{sma95}.  We kept as fixed
parameters in the fit $\sigma$, $\gamma$, and iterated on $m^{gc}_{peak}$,
as a function of the galaxy distance, while $N_{ogc}$ and $N_{obg}$ were
estimated by fitting the composite luminosity function to the external
sources extracted from the image in the range used for SBF
measurements in each galaxy.  Identified foreground stars were not
included in the fit.  From the estimated $N_{ogc}$ and
$N_{obg}$ per pixel, $P_{es}$ was calculated as the sum of
:
\begin{eqnarray}
\sigma^2_{gc}=&\frac{1}{2} N_{ogc} 10^{0.8[m_1-m^{gc}_{peak}+0.4\sigma^2ln(10)]}\\ \nonumber
& erfc[\frac{m_{cut}-m^{gc}_{peak}+0.8\sigma^2ln(10)}{\sqrt{2}\sigma}]
\end{eqnarray}
 and
\begin{equation}
\sigma^2_{bg}=\frac{N_{obg}}{(0.8-\gamma) ln(10)}10^{0.8(m_1-m_{cut})+\gamma(m_{cut})}.
\end{equation}
 \cite{bla95}, $m_{1}$ is the zero magnitude which corresponds to a
flux of 1 ADUs$^{-1}$.  The $P_{es}$ were calculated over the
luminosity function with this fitting procedure 
in each annulus, as in Sodemann \& Thomsen (1995).
To estimate $m_{cut}$ we have calculated the completeness function of
each annulus, adding artificial point sources to the original, galaxy
subtracted image.


\begin{figure}[!htf]

 \centering
\includegraphics[width=8cm]{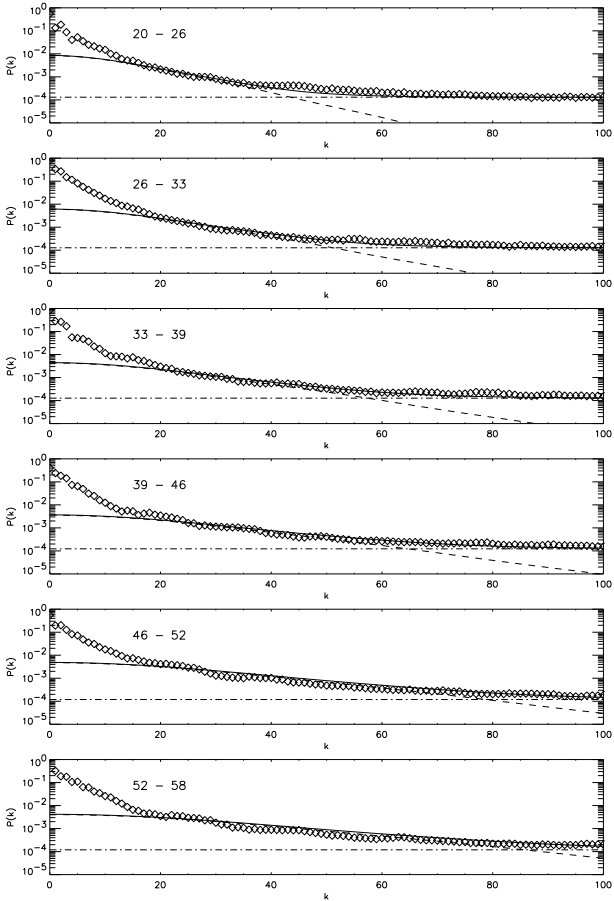}

\caption{NGC 7619 SBF measurement. We show NGC~7619 power spectrum as was fitted in six different annuli of width $\approx$ 6 $\arcsec$. The fit of the power spectrum is
given by the solid line, the PSF power spectrum by the dashed line
and the dashed-dotted line is the fitted constant white noise
spectrum. } \label{fig-7619spe}

\end{figure}


\begin{table*}[!htf]

\begin{flushleft}

\begin{tabular} {|c|c|c|c|c|c|c|c|c|c|c|c|c|c|} \hline 

Annulus ('') & P$_0$ & $\sigma_{P_0}$ & \small $P_{es}$&$\sigma_{P_es}$& (P$_0$ -- P$_{es}$)/P$_1$ & $(V-I)_o$ &$\overline M_I$&$\overline I_{o,k} $& $\sigma_{\overline I_{o,k}}$&$\overline I_{o,k} - \overline M_I$&$\sigma_{(\overline I_{o,k} - \overline M_I)}$\\ \hline
20 - 26 & 0.014 & 0.001&0.01 & 0.01&30&1.23& -1.38& 32.04 & 0.4& 33.42 &0.4\\ \hline
26 - 33&0.008&0.003&0.005&0.001&23&1.22&-1.43&32.16&0.4&33.58&0.4 \\ \hline
33 - 39&0.005&0.001&0.002&0.001&23&1.21&-1.47&32.34&0.4&33.81&0.4 \\ \hline
39 - 46&0.004&0.001&0.001&0.001&23&1.21&-1.47&32.26&0.3&33.73&0.4\\ \hline
46 - 52&0.005&0.001&0.0009&0.0005&34&1.21&-1.47&32.07&0.3&33.54&0.4\\ \hline
52 - 58&0.004&0.001&0.0009&0.0005&27&1.23&-1.38&32.39&0.3&33.77&0.4 \\ \hline

 \hline
 Mean  & -- &-- & --&--&--&--&--&--&--&33.64 & 0.16 \\ \hline

\end{tabular}

\end{flushleft}
\caption{SBF measurements for various annuli of NGC 7619.} \label{tab-7619annu}
\end{table*}


In the calculation of $P_{es}$ the fact that
$m_{cut}$ corresponds to a completeness function of $50\%$ has been
taken in account, integrating the contribution on the source completeness
function.

We show the respective external source luminosity functions of NGC~564 and NGC~7619 in Fig.~~\ref{fig-564lum} and Fig~~\ref{fig-7619lum}.

 The SBF amplitude is given by:
\begin{equation}
\overline m_I =  -2.5 log(P_0-P_{es}) + m_{1} - k_{ext} sec(z) 
\label{eq:mag}
\end{equation}
where $k_{ext}$ is the extinction coefficient and $z$ the airmass for the observations. In the I--band, we measured $k_{ext}=0.06 \pm 0.01$ and
$m_{1}=26.59 \pm 0.02$ for NGC~564, and $k_{ext}=0.07 \pm 0.01$ and
$m_{1}=26.60 \pm 0.02$ for NGC~7619. The effective airmass for
the nights that were used as photometric reference was $sec(z)$=1.2 
for NGC~564 and $sec(z)$=1.38 for NGC~7619, as calculated with the IRAF 
task {\it setairmass} on all images from that night (3 November 2000) 
for NGC~564, and on the photometric images used as a reference for NGC~7619.

In the V--band, we measured $k_{ext}=0.11 \pm 0.01$ for NGC~564 and $k_{ext}=0.12 \pm 0.01$ for NGC~7619, and $m_{1}=27.53 \pm 0.02$ for NGC~564 and
$m_{1}=27.45 \pm 0.02$ for NGC~7619. The effective airmass was $sec(z)$=1.2 for NGC~564 and  $sec(z)$=1.24 for NGC~7619.

\subsection[Results for NGC 564]{Results on NGC 564}

The power spectrum fitting of NGC~564 is shown in
Fig.~\ref{fig-564spe}. The results for each annulus are listed in
Table~\ref{tab-564annu}.  The SBF magnitudes we
calculated from Eq.~\ref{eq:mag}.
The errors on SBF magnitudes for each annulus
are the standard deviations among different wavelength cuts (3~$\lessapprox$ pixel scale $\lessapprox$~20). The
errors due to external source residual contribution subtraction were
added in quadrature to the fitting errors and to 
 the errors due to the zero point
magnitude and the $k_{ext}$.

 The SBF magnitudes $I_{o,k}$ were then corrected for
galactic absorption assuming $E(B-V) = 0.038$, $A_I=0.074$, and
$A_V=0.13$, from Schlegel et al. (1998), and a $k$--correction
$k_I\approx 7 \ z = 0.12$ for SBF was applied (Tonry et al. 1997; Liu et
al. 2000), $\overline I_{o,k}= \overline m_I - A_I - k_I $. The total
$A_I + k_I$ was equal to 0.19 mag.

 The galaxy colour in each annulus is shown in
Table~\ref{tab-564annu}. The error on each colour measurement was $0.03$ mag. The colours that are shown in the table have been corrected for
extinction. The adopted $A_I - A_V$ correction amounts to 0.056 mag. For the calculation of the galaxy colours the  $k$--correction is $k_I = 1 \ z $ and $k_V = 1.9 \ z $.

 From the Tonry et al. (2000) calibration :

\begin{equation}
\overline M_I=(-1.74 \pm 0.08)+(4.5 \pm 0.25) [(V-I)_0-1.15]. \label{eq:Ical}
\end{equation}

 we derive the $\overline M_I$ shown in Table~\ref{tab-564annu}. The error on each
value of $\overline M_I$ is $\sigma_{\overline M_I} = 0.18$. We did not add here the systematic uncertainty due to Cepheid calibration (Ferrarese et al. 2000a; Freedman et al. 2001)(see below).

From the calibration from Ferrarese et al. (2000a), revised to the Freedman et al. (2001) Cepheid distances (as from Ajhar et al. (2001)):

\begin{equation}
\overline M_I=(-1.73 \pm 0.09)+(4.5 \pm 0.25) [(V-I)_0-1.15]. \label{eq:Ical2}
\end{equation}

 we would have obtained a difference in SBF absolute magnitude of 0.01 mag and the same statistical errors. 

 A systematic uncertainty from the Cepheid calibration amounting to 0.16 mag has to be added to both calibration zero points. 

 In each annulus we derive an estimated distance modulus
$\overline I_{o,k} - \overline M_I$. A mean distance modulus was
determined as the mean of the values in the considered annuli. The
error is given by the standard deviation of those values, around the
mean, divided by the square root of the number of considered values.

 The final distance modulus is $(\overline I_{o,k} -
\overline M_I) = 34.39 \pm 0.18$ and the galaxy distance $76 \pm 6 $ Mpc.
If we use the SBF calibration by Ferrarese et al. 2000a, we obtain : $(\overline I_{o,k} -
\overline M_I) = 34.38 \pm 0.18$ and the galaxy distance $75 \pm 6 $ Mpc.

To the statistical error has to be added a systematic uncertainty due to the Cepheid calibration that amounts to $6$ Mpc.


\begin{table*}[!htf]

\begin{flushleft}
\begin{tabular} {cccc} \hline

Sample&Galaxy&$V_{it flow}$&$H_0$ \\ \hline
&&(km s$^{-1}$)&(km s$^{-1}$/Mpc) \\ \hline
Ferrarese et al. (2000a)&NGC 4881$^1$&7441&72.7 $\pm18.7$ \\
&NGC 4373$^2$&3118&85.9 $\pm$ 17.2 \\
&NGC 0708$^3$&4831&70.8 $\pm$ 8.6 \\ 
&NGC 5193$^3$&3468&67.3 $\pm$ 12.4 \\ 
 &IC 4296$^3$&3341&60.2 $\pm$ 11.2 \\ 
&NGC 7014$^3$&5041&75.2 $\pm$ 7.2 \\ \hline  

Mei et al. 2000&IC 4296&3341&67 $\pm$ 10 \\ \hline
This work & NGC 564&5208& 70 $\pm$ 7 \\ 
 & NGC 7619&3571& 68 $\pm$ 6\\ \hline \hline 

\end{tabular}

\end{flushleft}
$^1$ Thomsen et al. 1997 \\
$^2$ Pahre et al. 1999 \\
$^3$ Lauer et al. 1998  \\

\caption{SBF measurements of $H_0$} \label{tab-sbfho}
\end{table*}


\subsection[Results on NGC 7619]{Results on NGC 7619}

The power spectrum fitting of NGC~7619 is shown in
Fig.~\ref{fig-7619spe}. The results for each annulus are listed in
Table~\ref{tab-7619annu}. As for NGC~564, the errors on SBF magnitudes
for each annulus are the standard deviations among different
wavelength cuts. The errors due to external source residual
contribution subtraction were added in quadrature to the fitting
errors and to
the errors due to the zero point magnitude and the $k_{ext}$.

 The SBF magnitudes $I_{o,k}$ were then corrected for
galactic absorption assuming $E(B-V) = 0.079$, $A_I=0.153$, and
$A_V=0.261$, from Schlegel et al. (1998), and a $k$--correction
$k_I\approx 7 \ z = 0.08$ was applied (Tonry et al. 1997; Liu et
al. 2000), $\overline I_{o,k}= \overline m_I - A_I - k_I $. The total
$A_I + k_I$ was equal to 0.23 mag.

 The galaxy colour in each annulus is shown in
Table~\ref{tab-7619annu}. The error on each measurement is equal to 0.03 mag.
The colours that are shown in the table have been corrected for
extinction. The adopted $A_I - A_V$ correction amounts to 0.11  mag.

 From the Tonry et al. (2000) calibration we derive the $\overline M_I$ shown in Table~\ref{tab-7619annu}. The error on each
value of $\overline M_I$ is $\sigma_{\overline M_I} = 0.18$. As for NGC~564 
with the Ferrarese et al. (2000a) calibration, we would have obtained a difference in SBF absolute magnitude of 0.01 mag and the same statistical errors.
This value is compatible with Tonry et al. (2001) measurements ($\overline M_I = 33.62 \pm 0.31 $).

Again, a systematic uncertainty from the Cepheid calibration amounting to 0.16 mag has to be added to both calibration zero points. 

 In each annulus we derive an estimated distance modulus
$\overline I_{o,k} - \overline M_I$. A mean distance modulus was
determined as the mean of the values in the considered annuli. The
error is given by the standard deviation of those values, around the
mean, divided by the square root of the number of considered values.

 The final distance modulus is $(\overline I_{o,k} -
\overline M_I) = 33.64 \pm 0.16$ and the galaxy distance $54 \pm 4 $ Mpc.

\section[$H_0$ Measurement]{$H_0$ Measurement}

The value for $H_0$ is derived from our distance measurements and the
redshifts of the respective clusters of the galaxies. The cluster
redshift and Tully-Fisher peculiar velocity were used.

For NGC~564 the recession velocity of the cluster A194 is $cz_{CMB} =
5253 \pm 222~$km~s$^{-1}$ from Dale et al. (1999), and we derive the
value of $H_0$ = 70 $\pm$ 7 $\pm$ 6 km~s$^{-1}$/Mpc, where the second
error is the systematic uncertainty due to the Cepheid calibration.
For NGC~7619, in the Pegasus cluster, the cluster redshift is 3635
$\pm$ 221~km~s$^{-1}$ from Giovanelli et al. 1998, and we derive $H_0$
= 68 $\pm$ 6 $\pm$ 6 km~s$^{-1}$/Mpc.
When the Ferrarese et al. (2000a) calibration is used, we obtain the
same values for $H_0$.

We have also measured the distance of IC~4296 from previous VLT
 measurements (Mei et al. 2000). From that measurement (49 $\pm$ 4
 Mpc) and the redshift for IC~4296, and the flow corrected galaxy
 recession velocity $cz = 3341 \pm$ 552 km~s$^{-1}$ (Ferrarese et
 al. 2000b), $H_0$ = 68 $\pm$ 10 $\pm$ 6 km~s$^{-1}$/Mpc.

From our three galaxy sample, we obtain : $H_0$ = 68 $\pm$ 5 $\pm$ 6
km~s$^{-1}$/Mpc.

Our errors are comparable the {\it Hubble Space Telescope Key Project on the Extragalactic Scale}. We can then measure the Hubble constant from the Ferrarese et al. (2000a)  SBF sample (Freedman et al. 2001 Table 10) and our VLT three galaxy sample, calibrated on the Ferrarese et al. (2000a) calibration.

The two samples (Ferrarese et al. and our sample calibrated with the Ferrarese et al. calibration, that gives us the same results then the Tonry et al. calibration) are shown in Table~\ref{tab-sbfho}, where the statistical errors on $H_0$ are given. IC~4296 is present in both Ferrarese and our sample.
From the two combined samples we obtain  $H_0$ = 71 $\pm$ 4 $\pm$ 6 km~s$^{-1}$/Mpc. The first error is statistical, the second one is the systematic uncertainty due to the Cepheid calibration (Freedman et al. 2001).
Our results are comparable with other I--band and K--band SBF measurements, and, increasing the I--band SBF sample beyond 3000 km~s$^{-1}$,  we have reduced the statistical error on I--band SBF $H_0$ measurements from Freedman et al. 2001 from 5 to 4 km~s$^{-1}$/Mpc (their measurement is in fact: $H_0$ = 70 $\pm$ 5 $\pm$ 6 km~s$^{-1}$/Mpc).

If we combine only galaxies beyond 4000 km~s$^{-1}$ from the two samples,
 we obtain $H_0$ = 72 $\pm$ 6 $\pm$ 6 km~s$^{-1}$/Mpc).

There have been other two recent measurements of $H_0$ with I--band SBF, one from Tonry et al. 2001 with a sample of $\approx$ 300 ellipticals out to $\approx$~3000 km~s$^{-1}$, and their cosmic flow reconstruction($H_0$ = 77 $\pm$ 4 $\pm$ 7 km~s$^{-1}$/Mpc), and another one from Blakeslee et al. 2001 from the comparison of part of the Tonry sample ($\approx 160$ ellipticals) with Fundamental Plane distances and IRAS density  measurements  ($H_0$ = 73 $\pm$ 4 $\pm$ 7 km~s$^{-1}$/Mpc).

With K--band SBF measurements by NICMOS with HST, Jensen et al. 2001 have obtained $H_0$ = 72 $\pm$ 2.3 $\pm$ 6 km~s$^{-1}$/Mpc, obtaining for the first time in $H_0$ measurements by SBF a statistical error comparable to the supernovae Type Ia one (2 km~s$^{-1}$/Mpc (Freedman et al. 2001)). Their sample includes a dozen of galaxies with distances beyond 4000~km~s$^{-1}$.
Liu $\&$ Graham 2001 from Keck K--band SBF plus HST I--band  measurements have measured  $H_0$ = 71 $\pm$ 8 km~s$^{-1}$/Mpc.

All these measurements are compatible within the errors.

\section{Summary and Conclusions}

We have measured the Hubble constant $H_0$ in NGC~564 at $cz \approx
5800$ km~s$^{-1}$ and in NGC~7619 at $cz \approx 3700$ km~s$^{-1}$
with deep I--band SBF distance measurements at the ESO Very Large
Telescope (VLT). We obtain $H_0 = 70 \pm 7 \pm 6$ km~s$^{-1}$/Mpc for
NGC~564 and $H_0 = 68 \pm 6 \pm 6$ for NGC~7619. The actual SBF sample
used for the measurement of $H_0$ in the Hubble Space Telescope Key
Project on the Extragalactic Distance Scale (Freedman et al. 2001)
amounts to six galaxies.

When we combine the measurements from this work with our previous VLT
I--band SBF distance measurement in IC~4296 (Mei et al. 2000), we
obtain : $H_0$ = 68 $\pm$ 5 $\pm$ 6 km~s$^{-1}$/Mpc. When we add the
Freedman et al. (2001) SBF sample and calibrate our sample with the
Ferrarese et al. (2000a) calibration, we obtain $H_0 = 71 \pm 4 \pm 6$
km~s$^{-1}$/Mpc.
If we combine only galaxies beyond 4000~km~s$^{-1}$ from the two samples,
 we obtain $H_0$ = 72 $\pm$ 6 $\pm$ 6 km~s$^{-1}$/Mpc).

Our results are comparable with other I--band and K--band SBF
measurements.
More importantly we have increased the number of galaxies
in the I--band sample with redshifts larger than 3000~km s$^{-1}$,
 and thereby reduced the statistical error on I--band SBF $H_0$
measurements from the Freedman et al. (2001) value of 5~km~s$^{-1}$/Mpc to 
4~km~s$^{-1}$/Mpc (their measurement is in fact: $H_0$ = 70 $\pm$ 5 $\pm$ 6 km~s$^{-1}$/Mpc).

Few galaxies beyond the Fornax cluster have I--band SBF measurements and only 
five (included NGC~564 from this paper) have distances beyond 4000~km~s$^{-1}$.
This lack of distant galaxies is responsible for the fact that the 
statistical error on $H_0$ obtained with SBF measurements is still 
comparable to the systematic uncertainty due to the Cepheid calibration, which
is not the case for SNIa and TF methods (with a statistical error of, respectively,  2~km~s$^{-1}$/Mpc and 3~km~s$^{-1}$/Mpc (Freedman et al. 2001))
SBF being one of the most precise distance indicators, it is important
to further reduce the statistical uncertainties. To achieve this, additional
SBF measurements beyond 4000~km~s$^{-1}$ are needed.

\begin{acknowledgements}
S. Mei acknowledges support from the European Space Agency External Fellowship programme. We thank our referee, John Blakeslee for his comments useful for the improvement of this paper.

\end{acknowledgements}

\end{document}